# A Web server to locate periodicities in a sequence

**Claude Pasquier, Vassilis I. Promponas,**
**Nikos J. Varvayannis and Stavros J. Hamodrakas**

*Department of Biology, Division of Cell Biology and Biophysics*
*University of Athens, Athens 15701, Greece*

## Abstract

**Summary:** *FT is a tool written in C++, which implements the Fourier analysis method to locate periodicities in aminoacid or DNA sequences. It is provided for free public use on a WWW server with a Java interface.*

**Availability:** *The server address is http://o2.db.uoa.gr/FT*

**Contact:** *shamodr@atlas.uoa.gr*

Periodical patterns and tandem repeats of residues are often found in DNA and protein sequences. In DNA, locating such periodicities may reveal structural and functional characteristics of the molecule (e.g. existence of Z-DNA or protein coding regions). In proteins, their presence helps towards an understanding of the molecular structure of a fibrous/structural protein employing the principle of conformational equivalence and it may suggest ways of ultramolecular assembly for the formation of higher order structure. Characteristic examples are periodicities found in a number of sequences of fibrous proteins (e.g. tropomyosin, McLachlan and Stewart 1976; myosin, McLachlan, 1993; keratins, McLachlan,1978 and collagen, McLachlan, 1977).

Two basic methods were used in the past to manipulate sequences in order to locate exact or approximate tandem repeats or patterns: Fourier analysis and the study of internal homologies. Recently, other methods have also been developed: some based on Fourier transform theory (McLachlan, 1993; Cheever et al., 1991; Cornette et al., 1987; Viari et al., 1990; Lazovic, 1996; Veljkovic et al., 1985), others on mathematical methods like mutual information (Korotkov et al., 1997) or the theory of fractals (Voss, 1992).

## Text output

## Query data for:

`CCC4`

**Selected residues=(Y 1.0)**

**Appearance of the selected residues:**

```
              1         2         3         4
     12345678901234567890123456789012345678 90
0000 .Y........Y.........Y......Y.........Y
0040 .....Y......Y...........Y..........Y.
0080 .Y......Y.........Y......Y......
. . .
```

## Results

| n | a | b | r | theta | origin | period | intensity |
|---|---|---|---|---|---|---|---|
| ... | ... | ... | ... | ... | ... | ... | ... |
| 54 | 0.058 | −0.409 | 0.413 | −81.937 | 2.517 | 9.481 | 0.171 |
| 55 | 1.232 | −0.771 | 1.453 | −32.027 | 1.582 | 9.309 | 2.111 |
| 56 | 2.252 | 0.349 | 2.279 | 8.802 | 0.843 | 9.143 | 5.194 |
| 57 | 1.769 | 1.997 | 2.668 | 48.463 | 0.150 | 8.982 | 7.116 |
| 58 | 0.111 | 2.548 | 2.550 | 87.507 | −0.509 | 8.828 | 6.503 |
| 59 | −1.183 | 1.643 | 2.024 | 125.766 | −1.132 | 8.678 | 4.098 |
| 60 | −1.246 | 0.395 | 1.307 | 162.433 | −1.707 | 8.533 | 1.709 |
| ... | ... | ... | ... | ... | ... | ... | ... |

## Graphical output

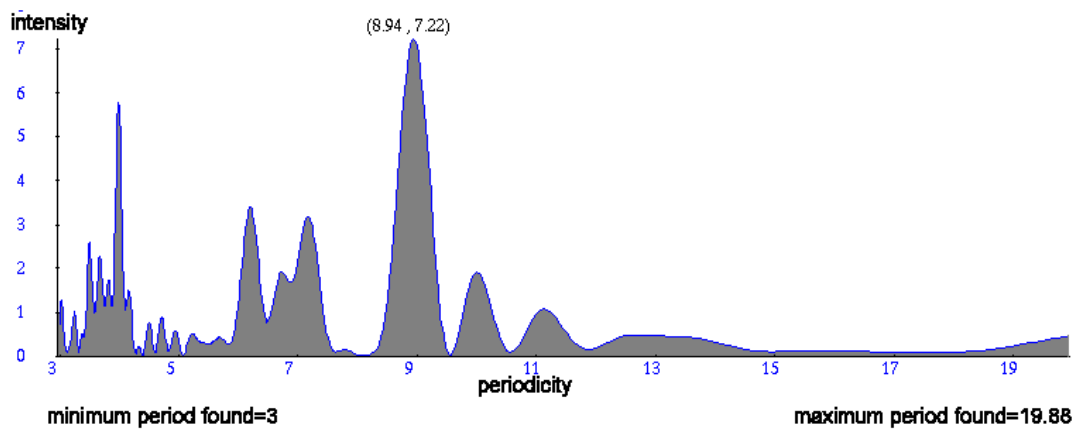

**Fig. 1.** A sample output produced by the program, after a search of periodicities for Tyrosine (Y) was made, between residues 162 and 278 of CCC4, a protein found in the eggshell of the fruit-fly *Ceratitis capitata* (Vlahou et al., 1997).

This application note describes in brief  FT, a tool, freely available through the Internet (URL 'http://o2.db.uoa.gr/FT'), which uses the  Fourier analysis method (McLachlan, 1977), to locate residue periodicities in aminoacid or DNA sequences.  The core program (which performs the Fourier transform) is written in C++. It can be executed on our machine (an o2 Silicon Graphics with a 180 MHz R5000 processor and 64 Mbytes of main memory) by disseminate users through the Internet.

Fourier transforms are obtained as outlined by McLachlan (1977). A sequence of N residues is represented as a linear array of N terms, with each term given a weight. The sequence of weights is used to create the pulse analysed by the program. For example, by selecting the weight 1 for 'A' and 2 for 'L', the sequence 'MISLIAALAVD' will be transformed by the program into the array {0 0 0 2 0 1 1 2 1 0 0}. The weight assigned to a residue could represent a special property of the residue (the charge, or the hydrophobicity for example) and can have a non-integer value. The use of a suitable combination of weights improves, significantly it seems, the detection of special characteristics of the sequence (Aggeli et al. 1991; Hamodrakas et al., 1985).

Directly on the form which appears on their web browsers, users can type or copy a sequence in one of three different formats (FASTA, SwissProt and PDB) and possibly select a part or the whole sequence for analysis. After selecting the residue or group of residues they wish to search for periodical appearance, they can run the program on our server and visualize the result on their web browser, usually in a few seconds. Results are presented in a table and can be displayed either as an HTML page or in text mode. On the HTML page, the relation between intensities and periodicities is also represented as a graph (Figure 1) .

To make the tool user-friendly, a search for periodic patterns of several groups of residues (α-helix formers, β–sheet formers, β–turn formers, hydrophobic, polar, charged, positively charged, negatively charged, aromatic, aliphatic) can be made simply by pressing a button. The manual of the package (freely available at URL 'http://o2.db.uoa.gr/FT/doc_index.html'), explains in detail how an association with available tables of properties of residues (e.g hydrophobicity scales of aminoacid residues) can be made easily. Also, since in most cases it is not known in advance if a residue will show a periodic pattern, the user can search for tandem repeats of all residues in a sequence simply by pressing a button. Most frequent residues, which, usually, have greater probability of showing periodic patterns, are treated first. The manual also contains some test cases for instructive purposes, with comments.

The data-input part of the program is written in Java. This allows users to benefit from an interface more user-friendly that those designed with pure HTML forms. The Java program controls the data entered and prints possible error messages before calling the core program. It also displays some interesting information concerning the data entered (length of the sequence, statistics on the appearance of each residue etc).

## Acknowledgments

The authors gratefully acknowledge the support of the EEC-TMR "GENEQUIZ" grant ERBFMRXCT960019.